\documentclass[useAMS,usenatbib]{mn2e}

\usepackage{graphicx}
\usepackage{hyperref}
\usepackage{txfonts}
\usepackage{upgreek}
\usepackage{multirow}
\usepackage{longtable}

\def\setsymbol#1#2{\expandafter\def\csname #1\endcsname{#2}}
\def\getsymbol#1{\csname #1\endcsname}

\newcommand{\arxiv}[1]{}

\newcommand{\arxivpreprint}[1]{preprint (\href{http://arxiv.org/abs/#1}{arXiv:#1})}

\def\aj{AJ}					
\def\apj{ApJ}					
\def\mnras{MNRAS}					
\def\apjs{ApJS}					
\def\apjl{ApJL}					
\def\aap{A\&A}					
\def\aaps{A\&AS}				
\def\memsai{Mem.~Soc.~Astron.~Italiana}		

\def\Planck{{\it Planck\/}}
\def\WMAP{{\it WMAP\/}}
\def\Herschel{{\it Herschel \/}}
\def\IRAS{{\it IRAS\/}}
\def\modot{\,M$_\odot$yr$^{-1}$}

\title{Radio to infrared spectra of late-type galaxies with \Planck~and \WMAP~data}
\author[M.\,W. Peel et al.]
 {M.\,W. Peel$^1$, C. Dickinson$^1$, R.\,D. Davies$^1$, D.\,L. Clements$^2$, R.\,J. Beswick$^1$\\
 $^1$ Jodrell Bank Centre for Astrophysics, Alan Turing Building, School of Physics and Astronomy, The University of Manchester,\\
$\phantom{^1}$ Oxford Road, Manchester, M13 9PL, U.K.\\
 $^2$ Imperial College London, Astrophysics group, Blackett Laboratory, Prince Consort Road, London, SW7 2AZ, U.K.
}
 
\begin{document}


\pagerange{\pageref{firstpage}--\pageref{lastpage}} \pubyear{2010}

\maketitle

\label{firstpage}

\begin{abstract}
We use the \Planck~Early Release Compact Source Catalogue combined with \WMAP~and other archival measurements to construct continuum spectra of three nearby dusty star-forming galaxies: Messier~82, NGC\,253 and NGC\,4945. We carry out a least-squares fit to the spectra using a combination of simple synchrotron, free-free and thermal dust models, and look for evidence of anomalous microwave emission (AME). We find that the radio spectra of all three galaxies are consistent with steep spectrum synchrotron emission, with a significant amount of free-free emission required to explain the \Planck~and \WMAP~data points in the frequency range 30-150\,GHz. This brings the star-formation rate based on free-free emission into better agreement with that from the non-thermal emission. We place limits on the presence of AME in these galaxies, finding that it is lower than expectations based on the ratio of far infrared to AME from the Galaxy. Nevertheless, the shape of the spectrum of NGC\,4945 hints at the presence of AME with a peak around 30\,GHz. Future \Planck~data will let us look more closely at these galaxies, as well as to extend the analysis to many more galaxies.
\end{abstract}


\maketitle

\section{Introduction}

\setsymbol{M82:Async}{$14.9\pm2.9$}
\setsymbol{M82:alphasync}{$-1.11\pm0.13$}
\setsymbol{M82:freefree}{$920\pm110$}
\setsymbol{M82:Adust}{$0.01\pm0.01$}
\setsymbol{M82:Betadust}{$2.10\pm0.13$}
\setsymbol{M82:Tdust}{$24.8\pm1.9$}
\setsymbol{M82:Chi2}{$0.86$}
\setsymbol{M82:CO100}{$0.32\pm0.14$}
\setsymbol{M82:CO217}{$0.78\pm0.17$}

\setsymbol{M82:residuals}{$-0.04\pm0.05$}
\setsymbol{M82:residuals:3sig}{$0.15$}

\setsymbol{M82:Asyncfixed}{$13.2\pm2.2$}
\setsymbol{M82:alphasyncfixed}{$-0.97\pm0.11$}
\setsymbol{M82:freefreefixed}{$770\pm130$}
\setsymbol{M82:Adustfixed}{$0.16\pm0.01$}
\setsymbol{M82:Tdustfixed}{$32.1\pm1.2$}
\setsymbol{M82:Chi2fixed}{$1.13$}

\setsymbol{NGC253:Async}{$11.1\pm4.3$}
\setsymbol{NGC253:alphasync}{$-1.59\pm0.35$}
\setsymbol{NGC253:freefree}{$284\pm17$}
\setsymbol{NGC253:Adust}{$0.12\pm0.07$}
\setsymbol{NGC253:Betadust}{$1.96\pm0.11$}
\setsymbol{NGC253:Tdust}{$22.6\pm1.3$}
\setsymbol{NGC253:Chi2}{$1.06$}
\setsymbol{NGC253:CO100}{$0.32\pm0.14$}
\setsymbol{M82:CO217}{$0.78\pm0.17$}

\setsymbol{NGC253:residuals}{$0.01\pm0.04$}
\setsymbol{NGC253:residuals:3sig}{$0.14$}

\setsymbol{NGC253:Asyncfixed}{$8.8\pm2.8$}
\setsymbol{NGC253:alphasyncfixed}{$-1.33\pm0.29$}
\setsymbol{NGC253:freefreefixed}{$263\pm22$}
\setsymbol{NGC253:Adustfixed}{$0.60\pm0.04$}
\setsymbol{NGC253:Tdustfixed}{$26.3\pm0.8$}
\setsymbol{NGC253:Chi2fixed}{$1.29$}

\setsymbol{NGC4945:Async}{$12.3\pm3.1$}
\setsymbol{NGC4945:alphasync}{$-1.15\pm0.20$}
\setsymbol{NGC4945:freefree}{$492\pm81$}
\setsymbol{NGC4945:Adust}{$0.004\pm0.004$}
\setsymbol{NGC4945:Betadust}{$2.5\pm0.2$}
\setsymbol{NGC4945:Tdust}{$18.9\pm1.1$}
\setsymbol{NGC4945:Chi2}{$0.92$}

\setsymbol{NGC4945:residuals}{$0.00\pm0.04$}
\setsymbol{NGC4945:residuals:3sig}{$0.13$}

\setsymbol{NGC4945:Asyncfixed}{$8.0\pm0.8$}
\setsymbol{NGC4945:alphasyncfixed}{$-0.63\pm0.03$}
\setsymbol{NGC4945:freefreefixed}{0}
\setsymbol{NGC4945:Adustfixed}{$0.40\pm0.02$}
\setsymbol{NGC4945:Tdustfixed}{$27.2\pm0.9$}
\setsymbol{NGC4945:Chi2fixed}{$1.91$}

The radio spectra of late morphological type nearby galaxies (Sb-Sc) have typically been measured at frequencies up to a few tens of Gigahertz, where the emission is predominantly synchrotron radiation. A limited number have been measured up to $\sim$100\,GHz, where the long wavelength tail of the thermal dust emission begins to contribute. The high frequency radio spectra of galaxies are important to identify optically thin free-free emission, which provides a measure of the star-formation rate (SFR), as well as identifying additional synchrotron components that result in flat synchrotron spectra at high frequencies, and constraining the amount of anomalous microwave emission (AME). \Planck~data provide the opportunity to identify AME in the range 20--60\,GHz~and free-free emission in the range 60--100\,GHz, where these mechanisms may dominate the spectrum of the galaxies, and can also detect or constrain flattening synchrotron emission.

In our Galaxy (type Sbc), free-free emission is a major contributor on the plane \citep{2010Alves} at 10--100\,GHz, but not at higher Galactic latitudes. Although free-free emission has been detected in spiral galaxies, it only becomes comparable to synchrotron emission at $\gtrsim$10\,GHz \citep{1997Niklas}. Hence we should expect \Planck~to detect free-free from nearby galaxies.

Diffuse AME, thought to be from spinning dust \citep[e.g.][]{1998Draine,2011PlanckAnom}, dominates the emission in our Galaxy both on the plane \citep{2010Alves} and at intermediate latitudes \citep{2003Banday,2003Lagache,2006Davies,2008Miville} at frequencies of 20--40\,GHz. AME is also present in a number of compact and diffuse regions at various latitudes (see e.g. \citealp{2005Watson,2011PlanckAnom}). However, to date, extragalactic AME has only been detected from some star-forming regions of the spiral galaxy NGC\,6946 \citep{2010Murphy,2010Scaife}.

We use the \Planck~Early Release Compact Source Catalogue (ERCSC) to look at three nearby dusty and star-forming galaxies. In $\S$2, we describe this catalogue and the ancillary data we combine with the catalogue. We then look at the individual spectra of three galaxies in $\S$3. Our conclusions are presented in $\S$4.

\section{Data}

The \Planck~ERCSC \citep{2011PlanckERCSC} covers nine frequency bands in the range 28.5--857\,GHz, using data from both the Low and High Frequency Instruments (LFI and HFI respectively) on board the \Planck~satellite. The catalogue covers the whole sky, and includes sources down to a few hundred mJy; however, it is not a complete flux density limited catalogue due to the source detection methodology and quality constraints. We have selected dusty and star-forming galaxies of late morphological type from the ERCSC, requiring that the galaxy has strong dust emission at 545\,GHz, be readily detectable by \Planck~at 28.5\,GHz, and that there are no strong contaminating compact sources within the \Planck~beams (which excludes NGC\,6503 due to the nearby quasar HB89 1749+701). We find 3 galaxies that satisfy these criteria: Messier~82 (Irregular), NGC\,253 (Sc) and NGC\,4945 (Sc).

To establish comprehensive SEDs, we combine measurements of these galaxies in the ERCSC with those in the Wilkinson Microwave Anisotropy Probe (\WMAP) 7-year point source catalogue \citep{2011Gold} and the Imperial \IRAS-FSC Redshift Catalogue \citep{2009Wang}. We also include other archival measurements by instruments that will not have resolved out a significant fraction of the emission from the galaxies.

Contamination from the Cosmic Microwave Background (CMB) anisotropies can be significant at frequencies $\sim\!100$\,GHz, particularly on scales of $\sim\!1^\circ$. The ERCSC flux densities should not contain significant amounts of CMB contamination as the extraction methods used will filter out emission on scales larger than the source extraction aperture. The source flux densities in the \WMAP~7-year main and CMB-subtracted catalogues are consistent with each other within their uncertainties (with the exception of the \WMAP~60.6\,GHz flux density for NGC\,4945; see later).

We conservatively add 5 per cent uncertainty \citep[following][]{1977Baars} to all archival measurements to account for the range of different calibration methods and beam sizes, and 13 per cent uncertainty to the \IRAS~measurements \citep{2005Miville}. We add 3 per cent uncertainty to the \WMAP~and \Planck~measurements, except for the \Planck~545 and 857\,GHz measurements where we add 7 per cent uncertainty \citep{2011PlanckHFI}. We apply colour corrections to the \Planck~and \WMAP~measurements (the latter using the fit provided by \citealp{2003Jarosik}) based on an estimate of the spectral index in each band.

We find that our results are robust to using either the `aperflux' or `gauflux' values from the ERCSC, which use 1.5 times the beam size as an aperture, and an elliptical Gaussian fit, respectively. For our analysis we use the ERCSC `aperflux' values for LFI (28.5--70.3\,GHz), where the large beam (13--33~arcmin) encompasses most of the extended emission from the galaxies, and `gauflux' for HFI (100-857\,GHz) as `aperflux' could miss some of the galaxy emission due to the smaller ($\sim$5~arcmin) beam size.

We perform a least-squares fit to the synchrotron, free-free and thermal dust components using:
\begin{equation}
S(\nu) = A_\mathrm{sync} \nu^\alpha + S_\mathrm{ff} + \frac{A_\mathrm{dust}h}{k} \frac{\nu^{\beta+3}}{\exp(h\nu/kT_\mathrm{dust})-1}~,
\end{equation}
i.e. we fit for the normalisation $A_\mathrm{sync}$ and spectral index $\alpha$ for the high frequency synchrotron radiation, and the normalisation $A_\mathrm{dust}$, spectral index $\beta$ and temperature $T_\mathrm{dust}$ of a grey body spectrum for the thermal dust. The free-free flux density is calculated by $S_\mathrm{ff} = 2\times 10^{26} \, k \, T_\mathrm{e}(1-e^{-\tau_\mathrm{ff}}) \, \Omega \, \nu^2 c^{-2}$, where $k$ is the Boltzmann constant, $T_\mathrm{e}$ is the electron temperature, which we set to the typical value for our Galaxy of 8000\,K, $\Omega$ is the solid angle of the galaxy (based on its optical size), and $\nu$ is the frequency. The optical depth $\tau_\mathrm{ff}$ is given by $\tau_\mathrm{ff} = 3.014 \times 10^{-2} \, T_\mathrm{e}^{-1.5} \, \nu^{-2} \,  {\rm EM_\mathrm{ff}} \, g_\mathrm{ff}$, where $g_\mathrm{ff}$ is the gaunt factor, which causes curvature in the high frequency free-free spectrum and is approximated as $g_\mathrm{ff} = \ln \left(4.955 \times 10^{-2} / \nu\right) + 1.5 \, \ln(T_\mathrm{e})$. We fit for the free-free emission measure, EM$_\mathrm{ff}$, in units of cm$^{-6}$\,pc over the extent of the optical emission. When calculating the residual values after subtraction of the model, we add the model uncertainties for synchrotron, free-free and thermal dust in quadrature with the measurement uncertainty.

Due to contributions from different emission mechanisms and physical parameters, the following data are not included in the fit:
\begin{enumerate}
\item Data from below 1.5\,GHz~where the synchrotron radiation might have a different, flatter, spectrum. This is because synchrotron spectra steepen at high frequencies due to synchrotron ageing (e.g. in our own Galaxy, see \citealp{2006Davies}). The lower frequencies are also affected by optically thick free-free absorption.
\item The 100 and 217\,GHz~HFI channels due to potential contamination by the $^{12}$CO(1--0) and $^{12}$CO(2--1) lines respectively (see \citealp{2011PlanckHFI}).
\item Far infrared data above 4\,THz to avoid small dust grain contributions to the large grain spectrum at lower frequencies.
\end{enumerate}

We use the residuals from these data after the subtraction of the model to set limits on the amount of AME from these galaxies. Note that the residual will contain a reduced amount of AME, as potential AME will have been fitted out by the free-free emission. We compare these with the values from AME regions within our Galaxy, where the ratio of AME at 30\,GHz~from HII regions and diffuse ionized gas to the 100\,$\upmu$m thermal dust flux density is $\sim3 \times 10^{-4}$ \citep{2010Todorovic,2010Alves} -- although this depends on the temperature, environment and grain population of the dust. The free-free and AME at 30\,GHz from the integrated emission through the Galactic plane (which is responsible for most of the total flux density from both emission mechanisms within our Galaxy) are comparable; this also would appear to be the expectation for the ionized gas in other galaxies similar to our own.

\begin{figure*}
\centering
\includegraphics[scale=0.23]{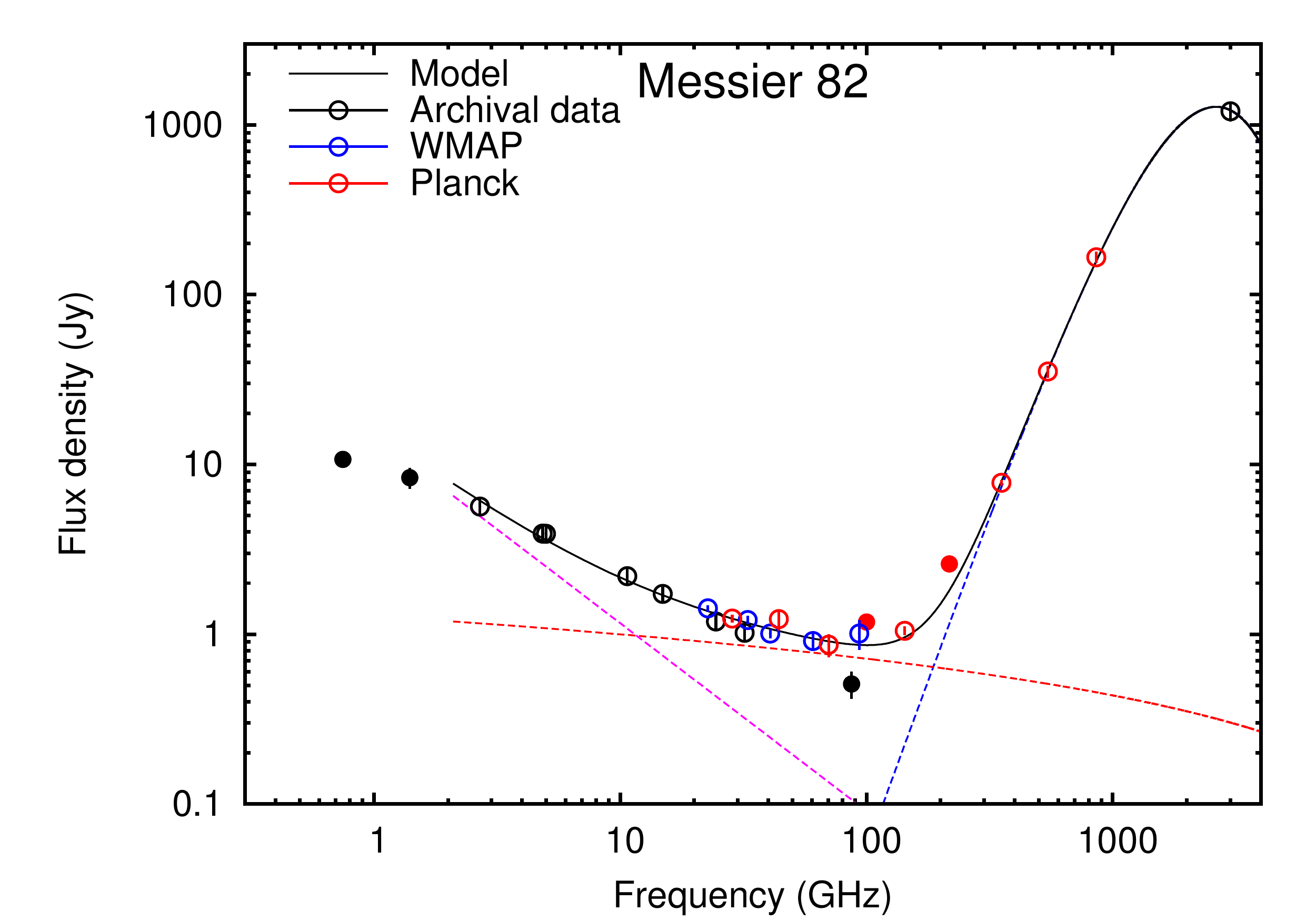}
\includegraphics[scale=0.23]{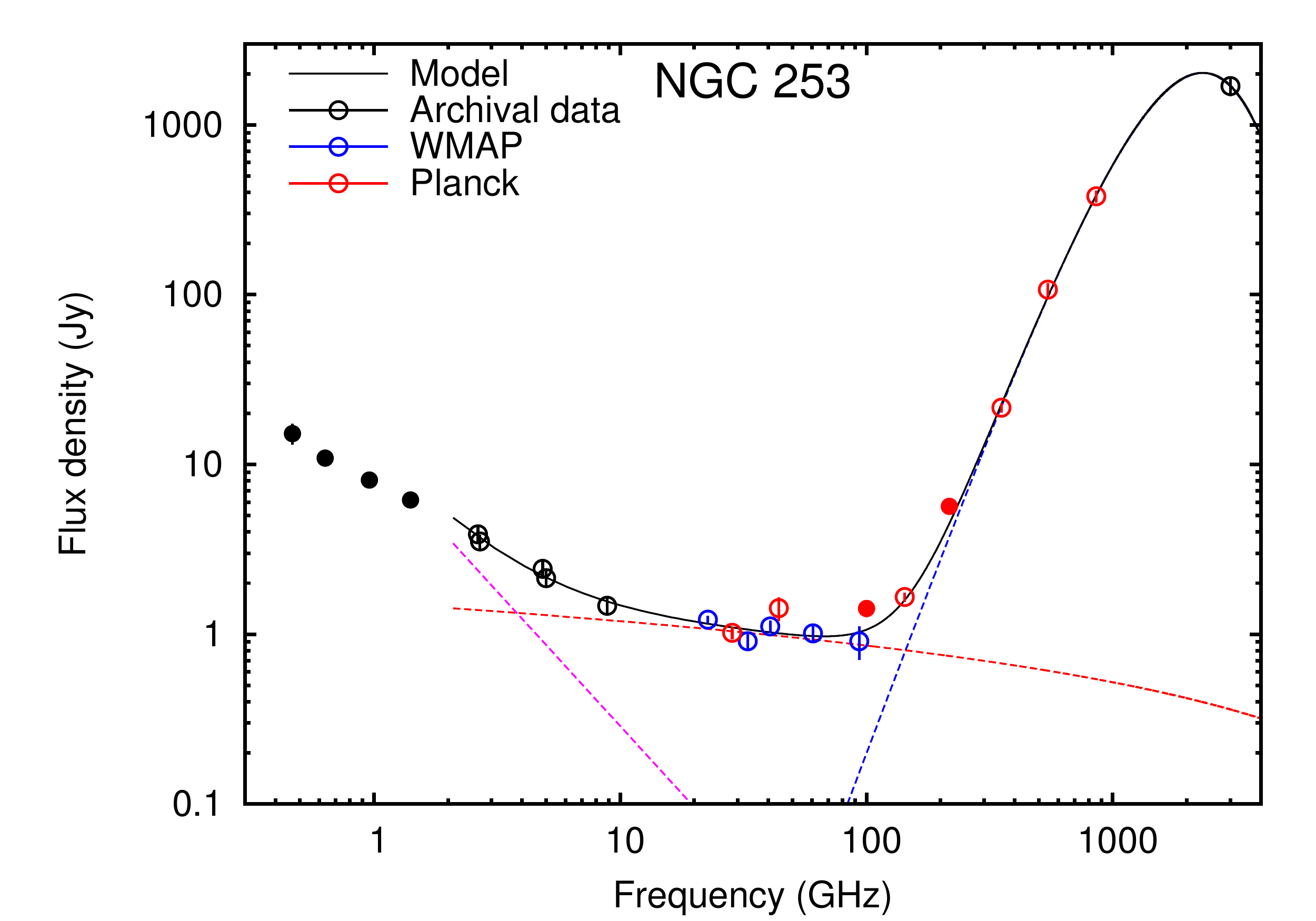}
\includegraphics[scale=0.23]{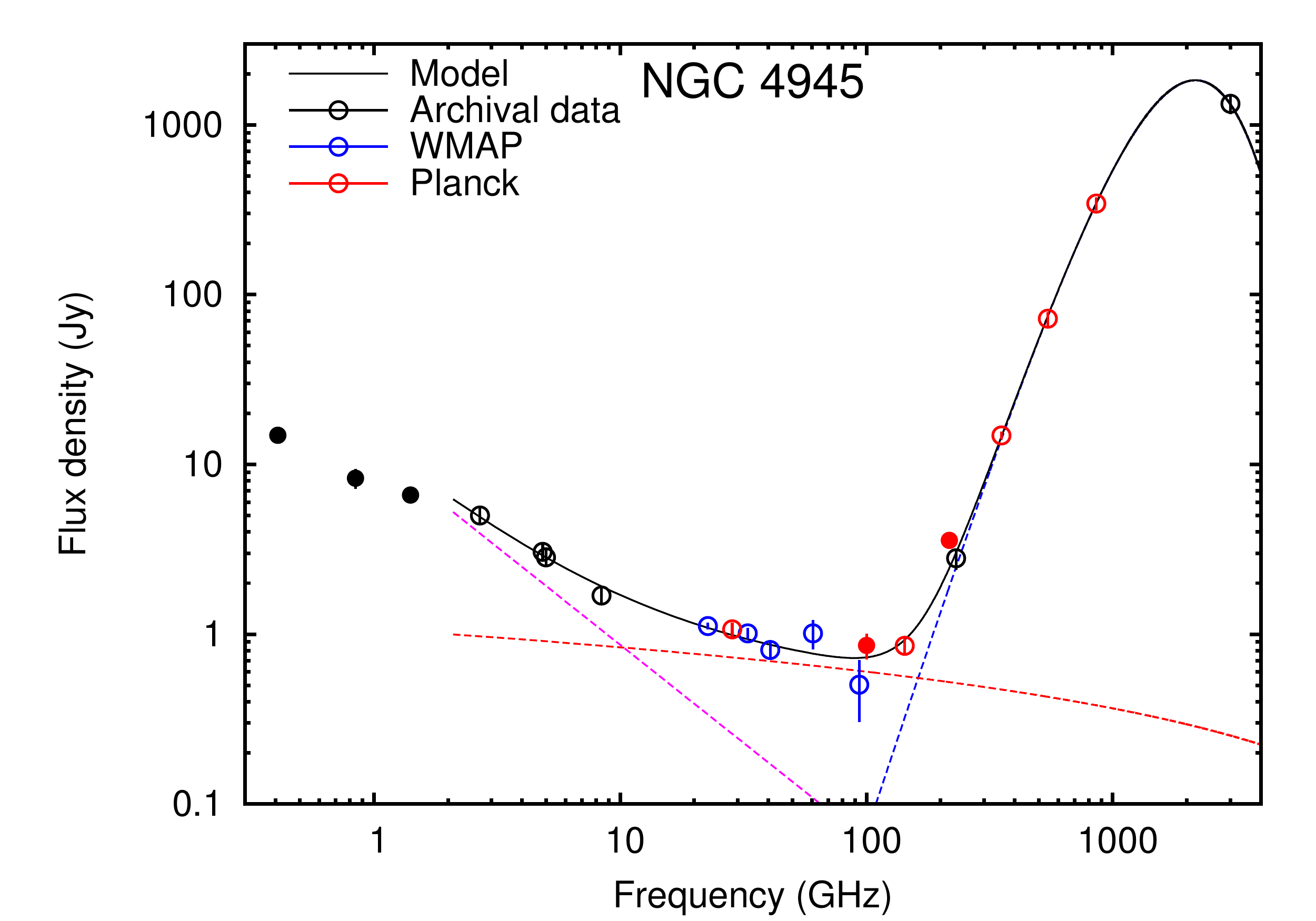}\\
\includegraphics[scale=0.23]{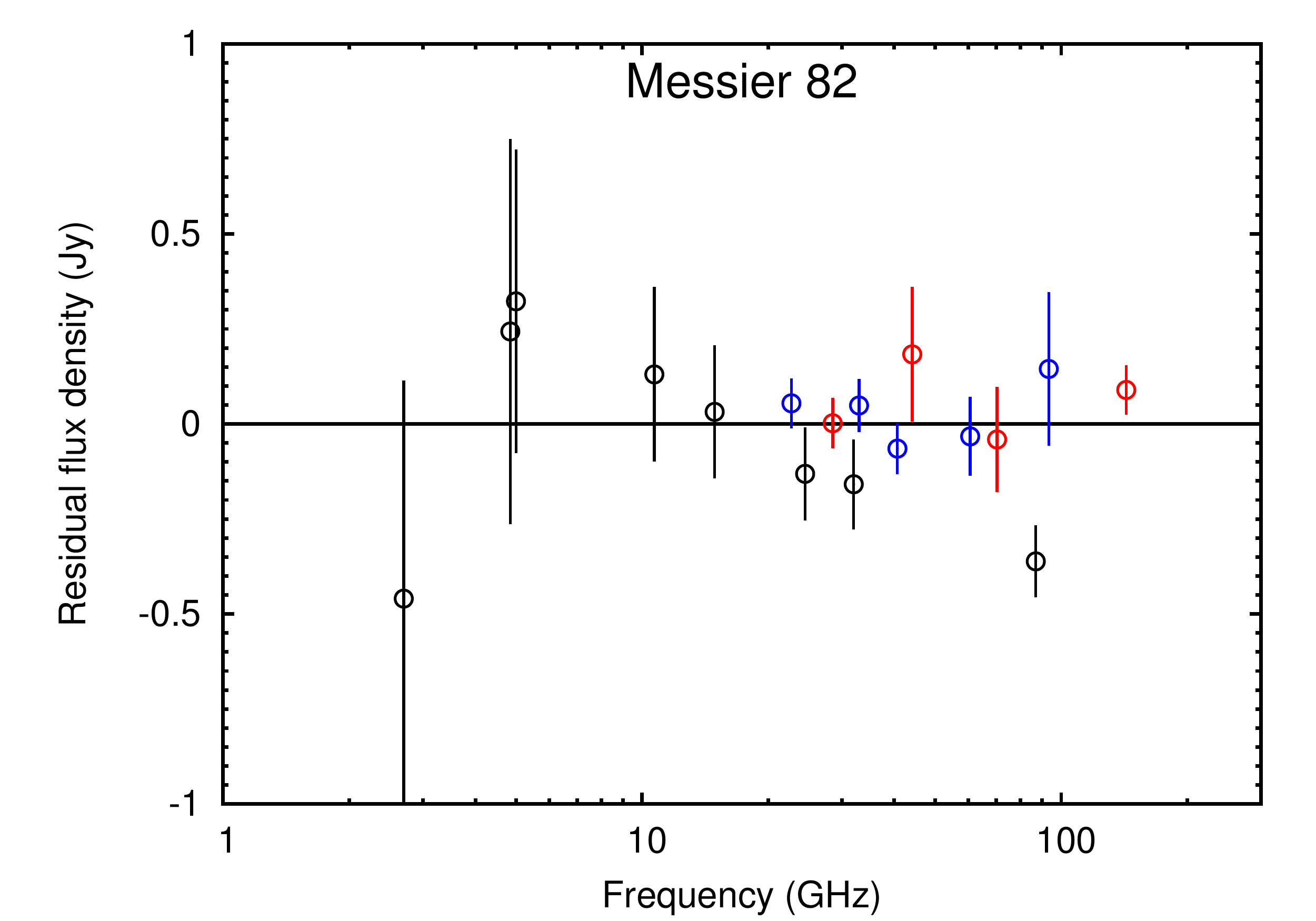}
\includegraphics[scale=0.23]{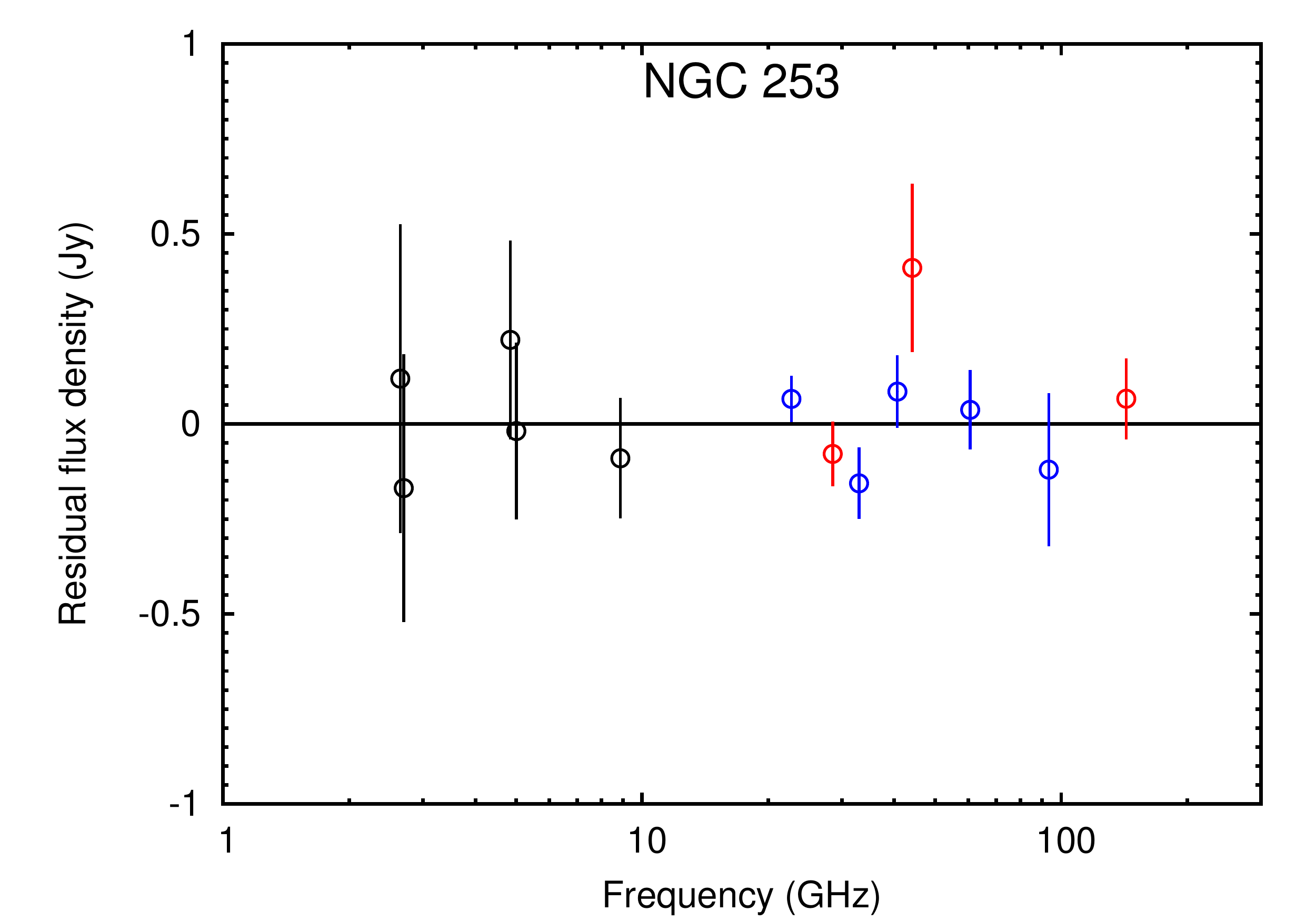}
\includegraphics[scale=0.23]{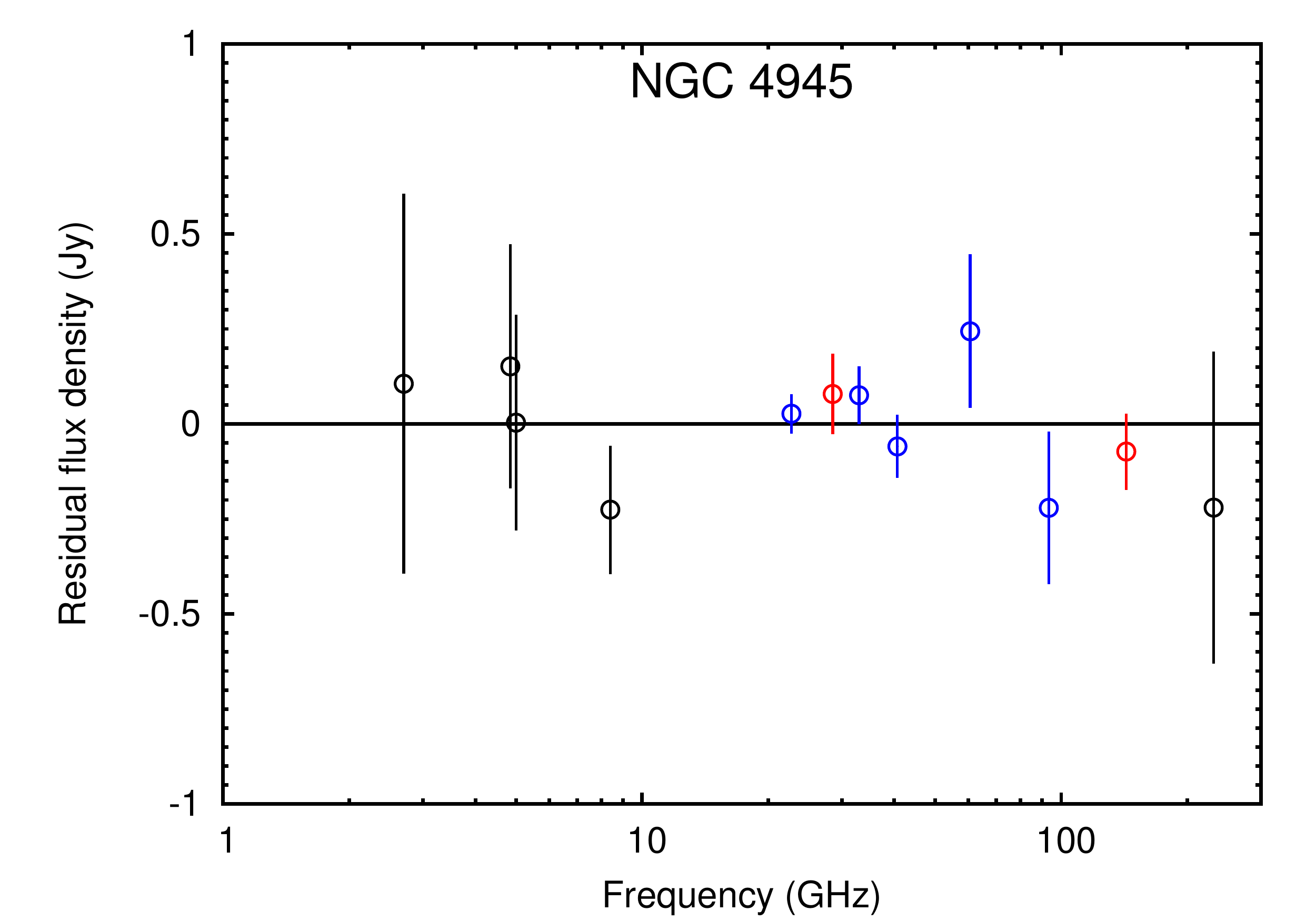}\\
\caption{{\bf Top}: Spectra for M82 (left), NGC\,253 (middle) and NGC\,4945 (right), including the best fit lines for synchrotron (magenta), free-free (red) and thermal dust (blue); the solid line is all three combined. Red data points are from the ERCSC, blue data points from the \WMAP~7-year catalogue and black data points are from NED (low frequency) and \IRAS~(high frequency). Solid points are not included in the least-squares fit. {\bf Bottom:} residual spectra for the same galaxies, excluding the CO-contaminated 100 and 217\,GHz data points. The error bars include synchrotron, free-free and thermal dust model fit uncertainties.}
\label{fig:spectra}
\end{figure*}

\section{Individual galaxies}
In the subsections below, we discuss the data sources used for each individual galaxy, the best fit model parameters that we find, and how these parameter values compare with others in the literature. The best-fitting parameters are summarised in Table 1. Our conclusions based on all three galaxies are presented in $\S$4.

\subsection{Messier~82}
Messier~82 ($09^\mathrm{h}55^\mathrm{m}52.7^\mathrm{s}, +69^\circ40^\mathrm{m}46^\mathrm{s}$ J2000.0), also known as NGC\,3034 and G141.40+40.56, is an irregular starburst galaxy at a distance of $3.3\pm0.5$~Mpc \citep{1988Freedman}. Its optical dimensions are $11.2\times4.3$ arcmin \citep{1991deVaucouleurs}, and at 857\,GHz~it is unresolved by \Planck~with a beam size of $5.3\times4.2$~arcmin.

We use low-frequency data from \citet{1980Laing, 1988Klein,1990Israel, 1991Gregory,1991Hales,1992White,1995Hales,2007Cohen}. We have excluded pre-1980 data as M82 has variability on long timescales due to supernovae outbursts, most notably 41.95+575 (see \citealp{2005Muxlow}). We note that there was another supernova in 2008 \citep{2010Brunthaler}, but as this is not evident in the \WMAP~year-by-year measurements we therefore assume that the change in total emission was negligible in these data due to their sampling times and high frequency.

These data and the least-squares fit are shown in the left-hand panel of Fig.\,\ref{fig:spectra}, with the model residuals directly below. The fitted dust temperature and spectral index are $T_\mathrm{dust} =$~\getsymbol{M82:Tdust}\,K and $\beta=+$\getsymbol{M82:Betadust}. This low temperature agrees with the statistical results in \citet{2011PlanckNearbyGals}, and the temperatures of $22.3-25.6$\,K found by recent \Herschel SPIRE observations of the outskirts of M82 \citep{2010Roussel}, although it is substantially lower than the standard values in the literature of $\sim$45\,K (e.g. \citealp{1980Telesco,1988Klein}). Note that fixing the spectral index to the average for normal-type galaxies of $\beta=1.65$ results in a temperature of \getsymbol{M82:Tdustfixed}\,K.  The HFI flux density measurements are significantly higher than others for M82 in the literature. For example, \citet{2000Thuma} measure $S_\mathrm{240\,GHz} = 1.91\pm0.13$\,Jy (after CO subtraction) from a $5.3\times4.5$\,arcmin map using the IRAM 30\,m telescope; our model fit (based on \Planck~HFI) predicts $\sim$2.3\,Jy at the same frequency ($\sim3\sigma$ higher). This discrepancy is likely due to the much larger beam size of \Planck, which will capture more of the extended emission.

We find a best-fit free-free emission measure of \getsymbol{M82:freefree}\,cm$^{-6}$\,pc (using the optical size). This is significantly higher than previous measurements: \citet{1997Niklas} limit the fraction of thermal emission at 10\,GHz~to less than 3 per cent (i.e. $S_\mathrm{thermal,10\,GHz} \lesssim 0.07$\,Jy), however we estimate that $\sim$50 per cent of the emission at 10\,GHz~is due to free-free emission (i.e. $\sim$1.0\,Jy of 2.1\,Jy). As a consequence of this, we find that the high ($>1.5$\,GHz) radio frequency synchrotron has a spectral index of $\alpha =$~\getsymbol{M82:alphasync}, which is steeper than that at lower frequencies, possibly due to synchrotron ageing. If the free-free contribution is set to zero, then the fitted synchrotron index becomes $\alpha =$~$-0.70\pm0.07$, which is in agreement with previous values \citep[e.g.][]{1988Klein}. This difference has arisen due to the extra information available from \Planck, which enables a better separation of the free-free and synchrotron emission, and the larger high frequency flux density values than others in the literature (e.g. \citet{1978Jura} measure only 60 per cent of the emission expected from this model at 87\,GHz).

We can compare the level of free-free emission with Radio Recombination Line (RRL) observations, as both of these emission mechanisms are generated by the same electron population. M82 was the first extragalactic source beyond the Magellanic clouds where RRLs were measured; \citet{1977Shaver} reported the detection of H166$\alpha$ using the Westerbork interferometer. More recent high-resolution measurements have mapped out the RRL emission from distinct regions within the galaxy \citep{2004Rodriguez-Rico}. Measurements of the H53$\alpha$ RRL using a single dish by \citet{1989Puxley,1991Carlstrom} find that the electron temperatures must be unfeasibly low unless the majority of the 0.55\,Jy they detected at 93\,GHz~is due to free-free emission. With our model, we find a total flux density of 0.87\,Jy at 93\,GHz, of which 0.72\,Jy is from free-free emission, in agreement with their calculation of an electron temperature of $T_\mathrm{e} > 5000$\,K; for comparison, this is a reasonable electron temperature (e.g. our own Galaxy has $T_\mathrm{e}\sim8000$\,K).

The SFR within M82 can be calculated from the non-thermal synchrotron radiation by assuming an initial mass function (IMF) and supernova rate; similarly, it can also be estimated from the free-free emission by assuming stellar models. Using the approximate formulae given by \citet{1992Condon}, we find a synchrotron SFR of $\sim$2.6\modot~based on the 1.4\,GHz flux density, and a free-free SFR of $\sim$3.0\modot~based on the best-fit free-free EM. This compares well with the SFR of $1.8-2$\modot~based on the radio supernova rate \citep{2008Fenech}, which is expected to underestimate the total SFR as it is only measuring the upper end of the IMF. These estimates are in much better agreement than the $<0.2$\modot~limit derived from the free-free the upper limit by \citet{1997Niklas}.

Based on the 100\,$\upmu$m flux density, we would expect to see $\sim$0.36\,Jy of AME at 30\,GHz. Between 23 and 143\,GHz~inclusive, the weighted mean of the residuals is \getsymbol{M82:residuals}\,Jy, or less than \getsymbol{M82:residuals:3sig}\,Jy at 3$\sigma$, indicating that there appears to be significantly less AME within M82 compared to our own Galaxy. Additionally, the SFR makes it unlikely that a considerable amount of AME emission could have been mixed into the free-free emission, making this a robust constraint on the AME level. Finally, AME within our Galaxy is only present in certain regions, due to environment and grain populations, which makes it likely that the overall level of AME from a galaxy will have a smaller ratio to infrared emission than that from individual regions (as seen by \citealp{2010Murphy} in NGC\,6946).

\begin{table}
\caption{Best fit parameters for synchrotron, free-free and dust emission in M82, NGC\,253 and NGC\,4945. R is the residual flux density from 23 to 143\,GHz after subtraction of the best-fit model, and represents a limit on the AME from each galaxy. Top: Fitting for $\beta_\mathrm{dust}$. Bottom: Fixing $\beta_\mathrm{dust}=1.65$.}
\begin{center}
\begin{tabular}{lccc}
\hline
Parameter & M82 & NGC253 & NGC4945\\
\hline
$A_\mathrm{sync}$ [Jy] & \getsymbol{M82:Async} & \getsymbol{NGC253:Async} & \getsymbol{NGC4945:Async} \\
$\alpha_\mathrm{sync}$ & \getsymbol{M82:alphasync} & \getsymbol{NGC253:alphasync} & \getsymbol{NGC4945:alphasync}\\
$\mathrm{EM}_\mathrm{ff}$[cm$^{-6}$pc] & \getsymbol{M82:freefree} & \getsymbol{NGC253:freefree} & \getsymbol{NGC4945:freefree}\\
$A_\mathrm{dust}$ & \getsymbol{M82:Adust} & \getsymbol{NGC253:Adust} & \getsymbol{NGC4945:Adust}\\
$\beta_\mathrm{dust}$ & \getsymbol{M82:Betadust} & \getsymbol{NGC253:Betadust} & \getsymbol{NGC4945:Betadust}\\
$T_\mathrm{dust}$ [K] & \getsymbol{M82:Tdust} & \getsymbol{NGC253:Tdust} & \getsymbol{NGC4945:Tdust}\\
$\chi^2 / \mathrm{dof}$ & \getsymbol{M82:Chi2} & \getsymbol{NGC253:Chi2} & \getsymbol{NGC4945:Chi2}\\
$R$ [Jy] & \getsymbol{M82:residuals} & \getsymbol{NGC253:residuals} & \getsymbol{NGC4945:residuals}\\
$R(3\sigma)$ [Jy] & $<$\getsymbol{M82:residuals:3sig} & $<$\getsymbol{NGC253:residuals:3sig} & $<$\getsymbol{NGC4945:residuals:3sig}\\
\hline
$A_\mathrm{sync}$ [Jy] & \getsymbol{M82:Asyncfixed} & \getsymbol{NGC253:Asyncfixed} & \getsymbol{NGC4945:Asyncfixed} \\
$\alpha_\mathrm{sync}$ & \getsymbol{M82:alphasyncfixed} & \getsymbol{NGC253:alphasyncfixed} & \getsymbol{NGC4945:alphasyncfixed}\\
$\mathrm{EM}_\mathrm{ff}$[cm$^{-6}$pc] & \getsymbol{M82:freefreefixed} & \getsymbol{NGC253:freefreefixed} & \getsymbol{NGC4945:freefreefixed}\\
$A_\mathrm{dust}$ & \getsymbol{M82:Adustfixed} & \getsymbol{NGC253:Adustfixed} & \getsymbol{NGC4945:Adustfixed}\\
$\beta_\mathrm{dust}$ & 1.65 & 1.65 & 1.65\\
$T_\mathrm{dust}$ [K] & \getsymbol{M82:Tdustfixed} & \getsymbol{NGC253:Tdustfixed} & \getsymbol{NGC4945:Tdustfixed}\\
$\chi^2 / \mathrm{dof}$ & \getsymbol{M82:Chi2fixed} & \getsymbol{NGC253:Chi2fixed} & \getsymbol{NGC4945:Chi2fixed}\\
\hline
\end{tabular}
\end{center}
\label{default}
\end{table}

\subsection{NGC\,253}
NGC\,253 ($00^\mathrm{h}47^\mathrm{m}33.1^\mathrm{s}, -25^\circ17^\mathrm{m}18^\mathrm{s}$ J2000.0), also G097.18-87.96, is an Sc galaxy at a distance of 2.6$\pm$0.7\,Mpc \citep{1988PucheI} with a starburst in its nucleus. Its optical dimensions are $27.5\times6.8$~arcmin \citep{1991deVaucouleurs} and at 857\,GHz~its dimensions are $6.1\times5.0$\,arcmin (partially resolved).

We use low-frequency data from \citet{1973Shimmins,1975Wills,1981Kuhr,1990Israel, 1994Griffith}. NGC\,253 is not included in the 70.3\,GHz band of ERCSC. These data and the least-squares fit are shown in the central panel of Fig.\,\ref{fig:spectra}. The dust is again cold, with $T_\mathrm{dust}=$~\getsymbol{NGC253:Tdust}\,K and $\beta =$~\getsymbol{NGC253:Betadust} (or \getsymbol{NGC253:Tdustfixed}\,K fixing $\beta=1.65$). There is a significant amount of CO emission from NGC\,253 within various HFI bands (see \citealp{2011PlanckNearbyGals}), which should be kept in mind when considering the dust fit presented here. However, this will negligibly affect the results at lower frequencies, and does not affect the LFI or 143\,GHz~bands where there are no CO lines.

We find that the best fit value for the free-free EM is \getsymbol{NGC253:freefree}\,cm$^{-6}$\,pc, with a synchrotron spectral index of $\alpha=$~\getsymbol{NGC253:alphasync}, which is steep compared to lower frequencies. Without free-free, the synchrotron index has an index above 1.5\,GHz of $\alpha =$~$-0.77\pm0.11$. The H102$\alpha$ RRL in NGC\,253 was the second extragalactic RRL to be detected, and was observed by \citet{1977Seaquist} at 6.1\,GHz~with a flux density of $20\pm3$\,mJy and width of $309\pm65$\,km\,s$^{-1}$. At 6.1\,GHz, our model predicts 1.27\,Jy of free-free emission out of a total of 1.89\,Jy, which agrees with the low $T_\mathrm{e}$ ($<$3800\,K) found by \citet{1977Seaquist}.

NGC\,253 was included in the study by \citet{1997Niklas}, who found the fraction of thermal emission at 10\,GHz~to be 0.35 ($\sim0.55$\,Jy). We find a significantly higher fraction of free-free emission at 10\,GHz of 0.81 (1.2\,Jy) in our best-fit model, which is due to the higher frequency data from \Planck~and \WMAP. Based on the non-thermal radiation at 1.4\,GHz, the SFR is 1.3\modot; using the \citet{1997Niklas} value gives a thermal SFR of 1\modot, and using our model we find a thermal SFR of 2.2\modot, indicating that we are possibly overestimating the level of free-free emission.

For NGC\,253, the expected 30\,GHz AME based on the 100\,$\upmu$m flux density would be $\sim$0.5\,Jy. The weighted mean between 23 and 94\,GHz~of the residual is \getsymbol{NGC253:residuals}\,Jy, giving a $3\sigma$ upper limit of \getsymbol{NGC253:residuals:3sig}\,Jy. Again, this is significantly below that expected based on our Galaxy; however, based on the high free-free SFR from our model it is possible that AME has been combined with the free-free estimate; if around half of the free-free emission is actually AME then this would be a similar ratio as to our Galaxy.

\subsection{NGC\,4945}

NGC\,4945 ($13^\mathrm{h}05^\mathrm{m}27.5^\mathrm{s}, -49^\circ28^\mathrm{m}06^\mathrm{s}$ J2000.0), also known as G305.27+13.34, is an Sc galaxy at a distance of $\sim3.6$\,Mpc \citep{2002Karachentsev} with optical dimensions of $19.9\times3.8$~arcmin \citep{1991deVaucouleurs} and observed 857\,GHz~ dimensions of $6.82\times3.98$\,arcmin (partially resolved).

We use low frequency data from \citet{1990Parkes,1992Jones, 1994Wright, 2004Wang}. We use \IRAS~values from the catalogue by \citet{2003Sanders} as NGC\,4945 is not included in that by \citet{2009Wang}. NGC\,4945 is also not included in either the 44.1 or 70.3\,GHz bands of ERCSC (most likely due to the high signal to noise photometric accuracy requirement of the ERCSC). The right-hand panel of Fig.\,\ref{fig:spectra} shows these data and least-squares fit.

We find a free-free EM of \getsymbol{NGC4945:freefree}\,cm$^{-6}$\,pc with a synchrotron index of $\alpha = $~\getsymbol{NGC4945:alphasync}, once more steeper than lower frequencies (without considering free-free, the best fit synchrotron index is still steep at $\alpha = -0.95\pm0.12$). The dust is the coldest of the three galaxies, with $T_\mathrm{dust}=$~\getsymbol{NGC4945:Tdust}\,K and $\beta=$~\getsymbol{NGC4945:Betadust}. Fixing $\beta=1.65$ results in $T_\mathrm{dust}=~$\getsymbol{NGC4945:Tdustfixed}\,K, however this significantly over-predicts the emission at 143\,GHz, requiring no contribution to the spectra from free-free emission and hence a much flatter synchrotron index that does not fit these low frequency data.

The calculated SFR based on the non-thermal flux density at 1.4\,GHz is 2.7\modot; using the free-free model fitted here returns a thermal SFR of 2.9\modot, i.e. consistent within the uncertainty. RRL from the nuclear region of NGC\,4945 have recently been measured for the first time by \citet{2010Roy}, who detected the H92$\alpha$ line using ATCA and find that NGC\,4945 is the strongest known extragalactic RRL emitter, and is compatible with a model using $T_\mathrm{e}=5000$\,K. Unfortunately for our purposes, this observation has much higher angular resolution (3~arcsec) than the \Planck~measurements, so a direct comparison of the expected free-free levels is not possible. However, the estimate of the SFR based on the RRL is $2-8$\modot -- compatible with, but higher than, our estimates here.

For NGC\,4945, the expected 30\,GHz AME based on the 100\,$\upmu$m flux density would be at the level of $\sim$0.4\,Jy. Between 23 and 94\,GHz, the weighted mean of the residuals is \getsymbol{NGC4945:residuals}\,Jy, giving a $3\sigma$ upper limit of \getsymbol{NGC4945:residuals:3sig}\,Jy. However, the 23-41\,GHz~measurements of NGC\,4945 suggest the possibility of a 30\,GHz-peaked curve (the high 60.6\,GHz data point from \WMAP~is likely due to CMB contamination: the entry in the CMB-subtracted \WMAP~catalogue is $0.5\pm0.2$\,Jy compared with $1.0\pm0.2$\,Jy from the main catalogue). Including a spinning dust model in the fit gives $N_\mathrm{H}=(2.7\pm1.6)\times10^{16}$\,cm$^{-2}$, reducing the free-free emission to $394\pm74$\,cm$^{-6}$pc, with $\chi^2/\mathrm{dof}=0.79$. At the same time, the synchrotron index is steepened to $\alpha=-1.2\pm0.2$ with $A_\mathrm{sync}=14.5\pm3.7$\,Jy, and the dust has $T_\mathrm{dust}=19.5\pm1.2$\,K and $\beta=2.4\pm0.2$. In order to pin down the potential AME in this source, further data from \Planck~would be useful -- particularly the 44 and 70\,GHz data points that are not in the ERCSC. More recent and accurate measurements of the flux density of the entire galaxy at frequencies less than 20\,GHz would also shed further light on this potential AME.

\section{Conclusions}

We find that the three dusty galaxies examined here (Messier\,82, NGC\,253 and NGC\,4945) are well-fitted by a combination of synchrotron, free-free and thermal dust emission, with $\chi^2\sim1$ (see Table 1). The excess emission at 10\,GHz~proposed as free-free by \citet{1997Niklas} for the large sample of late-type galaxies is confirmed in the present study. Furthermore, we typically find a higher fraction of the 10\,GHz emission to be free-free compared with \citet{1997Niklas}. Free-free emission is a reliable estimator of SFR in late-type galaxies when supplemented by RRL to provide information on the electron temperature \citep[see][]{2010Murphya}; we estimate the SFR using the free-free emission, and find our estimates to be consistent with that from the non-thermal emission.

We find steep synchrotron spectra at high radio frequencies (comparable to our own Galaxy) due to the significant amount of free-free emission that is required to explain the higher frequency \WMAP~and \Planck~measurements. We also find that the thermal dust is best fitted with colder temperatures than the standard values in the literature, in agreement with \citet{2011PlanckNearbyGals}.

We place limits on the AME from these galaxies, although some AME will have been identified as free-free emission. We note that quantifying any potential AME is difficult due to the combination of different emission mechanisms and physical environments present within each galaxy and the degeneracies between each of these parameters. In particular, the synchrotron spectral index, the level of free-free emission, and the level of AME are highly correlated. Extra information (e.g. morphological differences or spectral line information) needs to be included within the analysis to better disentangle the emission mechanisms.

Further ancillary data could lead to a firmer measurement of AME and free-free emission from the galaxies studied here. Later compact source catalogues from \Planck, as well as improved low frequency measurements, will greatly reduce the uncertainties presented in this letter. Additionally, this analysis can be extended to more late type galaxies (e.g. Arp\,220) when measurements of weaker sources are available from the low frequencies of \Planck.

\section*{Acknowledgements}
This research is based on observations obtained with \Planck, an ESA science mission with instruments and contributions directly funded by ESA Member States, NASA, and Canada. This research has made use of the NASA/IPAC Extragalactic Database (NED) which is operated by the Jet Propulsion Laboratory, California Institute of Technology, under contract with the National Aeronautics and Space Administration. CD acknowledges an STFC Advanced Fellowship and an ERC grant under the FP7.

\label{lastpage}

\end{document}